\documentclass[aoas]{imsart}\usepackage[]{graphicx}\usepackage[]{color}
\makeatletter
\def\maxwidth{ %
  \ifdim\Gin@nat@width>\linewidth
    \linewidth
  \else
    \Gin@nat@width
  \fi
}
\makeatother

\definecolor{fgcolor}{rgb}{0.345, 0.345, 0.345}

\usepackage{framed}
\makeatletter
 {\par\unskip\endMakeFramed%
 \at@end@of@kframe}
\makeatother

\definecolor{shadecolor}{rgb}{.97, .97, .97}
\definecolor{messagecolor}{rgb}{0, 0, 0}
\definecolor{warningcolor}{rgb}{1, 0, 1}
\definecolor{errorcolor}{rgb}{1, 0, 0}
\newenvironment{knitrout}{}{} 

\usepackage{alltt}

\RequirePackage{amsthm,amsmath,amsfonts,amssymb}
\RequirePackage[authoryear]{natbib}
\RequirePackage[colorlinks,citecolor=blue,urlcolor=blue]{hyperref}
\RequirePackage{graphicx}

\startlocaldefs
\usepackage{booktabs}
\usepackage{color}
\usepackage{todonotes}
\usepackage{xifthen}
\usepackage{array}
\usepackage{ulem}
\usepackage{url}
\usepackage{txfonts}

\newcommand{\blanco}[1]{ } 

 \newcommand{\hl}{\textcolor{black}}
 \newcommand{\soutr}[1]{} 

\newcommand{\dmin}{d_{\tiny \mbox{min}}}

\newcommand{\zrmin}{z_r^{\tiny \mbox{min}}}

\newcommand{\dinfty}{d_{\infty}}

\newcommand{\latin}[1]{\textit{#1}}

\newcommand{\abk}[1]{\mbox{#1}\xdot}
\DeclareRobustCommand\xdot{\futurelet\token\Xdot}
\def\Xdot{%
  \ifx\token\bgroup.%
  \else\ifx\token\egroup.%
  \else\ifx\token\/.%
  \else\ifx\token\ .%
  \else\ifx\token!.%
  \else\ifx\token,.%
  \else\ifx\token:.%
  \else\ifx\token;.%
  \else\ifx\token?.%
  \else\ifx\token/.%
  \else\ifx\token'.%
  \else\ifx\token).%
  \else\ifx\token-.%
  \else\ifx\token+.%
  \else\ifx\token~.%
  \else\ifx\token.%
  \else.\ %
  \fi\fi\fi\fi\fi\fi\fi\fi\fi\fi\fi\fi\fi\fi\fi\fi%
}

\newcommand{\eg}{\abk{\latin{e.\,g}}}
\newcommand{\ie}{\abk{\latin{i.\,e}}}

\DeclareMathOperator{\Nor}{N} 
\DeclareMathOperator{\Var}{Var} 
\DeclareMathOperator{\sign}{sign} 
\renewcommand{\P}{\operatorname{\mathsf{Pr}}} 

\newcommand{\given}{\,\vert\,} 

\endlocaldefs
\IfFileExists{upquote.sty}{\usepackage{upquote}}{}
\begin{document}

\begin{frontmatter}
\title{The assessment of replication success \\ based on relative effect size}
\runtitle{The assessment of replication success based on relative effect size}

\begin{aug}
\author{\fnms{Leonhard} \snm{Held}\ead[label=e1]{leonhard.held@uzh.ch}},
\author{\fnms{Charlotte} \snm{Micheloud}\ead[label=e2]{charlotte.micheloud@uzh.ch}}
\and
\author{\fnms{Samuel} \snm{Pawel}\ead[label=e3]{samuel.pawel@uzh.ch}}
\address{
  Epidemiology, Biostatistics
  and Prevention Institute, 
  Center for Reproducible Science,
  University of Zurich,
  \printead{e1,e2,e3}}
\end{aug}
\begin{abstract}
Replication studies are increasingly conducted \hl{in order} to
confirm original findings. However, there is no established standard
how to assess replication success and in practice many different
approaches are used. The purpose of this paper is to refine and
extend a recently proposed reverse-Bayes approach for the analysis
of replication studies.  We show how this method is
directly related to the relative effect size, the ratio of the
replication to the original effect estimate. This perspective leads
to a new proposal to recalibrate
the assessment of replication success, the golden level. 
\hl{The recalibration ensures that for borderline significant original 
studies replication success can only be achieved if the replication effect 
estimate is larger than the original one. Conditional power for replication success
can then take any desired value if the original study is significant and the replication
  sample size is large enough.} 
Compared to the standard approach to require statistical  
significance of both the original and replication study, replication
success at the golden level offers uniform gains in project power
and controls the Type-I error rate if the replication sample
size is not smaller than the original one.  
An application to
data from four large replication projects shows that the \hl{new} 
approach leads to more appropriate inferences, as it
penalizes shrinkage of the replication estimate compared to the
original one, while
ensuring that both effect estimates are sufficiently convincing on their own. 
\end{abstract}


\end{frontmatter}


\section{Introduction}
Replication studies are conducted in order to investigate whether an
original finding can be confirmed in an independent study. 
Although replication has long been a central part of the scientific method in 
many fields, the so-called replication crisis 
\citep{Ioannidis2005, Begley2015} has led to increased interest in replication 
over the last decade. These developments eventually culminated in large-scale
replication projects that were conducted in various fields 
\citep{Errington2014, Klein2014, OSC2015, Ebersole2016, Camerer2016, Camerer2018, Cova2018, Klein2018}. 

Declaring a replication as successful is, however, not a
straightforward task, and currently used approaches include
\hl{statistical} significance of both the original and replication
studies, compatibility of their effect estimates, and meta-analysis of
the effect estimates. \hl{Many of the replication projects listed above also
  report the relative effect size, the ratio of the
  replication to the original effect estimate. For example, in \citet{Camerer2018} 
  the replication effect estimates were only half as large as the original ones
  on average and even smaller in \citet{OSC2015}. This gives
  clear evidence of a systematic bias of the original studies and
  strongly suggests that the original and replication study should not
  be treated as exchangeable. However, all the approaches mentioned
  above will give the same results if the order of studies would be reversed. }

In order to address \hl{this problem}, a new method has recently been
proposed in \citet{held2020}. The approach combines the analysis of
credibility \citep{matthews:2001,matthews:2001b} with a prior-data
conflict assessment \citep{box:1980}. Replication success is declared if
the replication study is in conflict with a sceptical prior that would
make the original study non-significant. \hl{This approach penalizes
  small relative effect sizes as we will see in more detail in the following.}

To introduce some notation, let $z_o = \hat \theta_o/\sigma_o$ and
$z_r = \hat \theta_r/\sigma_r$ denote the $z$-statistic of the
original and replication study, respectively. Here $\hat \theta_o$ and
$\hat \theta_r$ are the corresponding effect estimates (assumed to be 
normally distributed) of the unknown 
effect $\theta$ with standard errors $\sigma_o$ and $\sigma_r$, respectively. 
The corresponding one-sided 
$p$-values are denoted by $p_o=1-\Phi(z_o)$ and $p_r=1-\Phi(z_r)$, 
respectively, where $\Phi(\cdot)$ denotes the standard normal cumulative
distribution function. 
Let $c = \sigma_o^2/\sigma_r^2$ denote the variance
ratio of the squared standard errors of the original and replication
effect estimates. The squared standard errors are usually inversely
proportional to the sample size of each study, \ie
$\sigma_o^2 = \kappa^2/n_o$ and $\sigma_r^2 = \kappa^2/n_r$ for some
unit variance $\kappa^2$.  The variance ratio $c$ can then be
identified as the relative sample size $c=n_r/n_o$. The relative effect size
\begin{equation}\label{eq:d}
  d = \frac{\hat \theta_r}{\hat \theta_o} = \frac{1}{\sqrt{c}} \frac{z_r}{z_o}
\end{equation}
quantifies the size of the replication effect estimate $\hat \theta_r$ relative 
to the original effect estimate $\hat \theta_o$. The corresponding shrinkage
of the replication effect estimate will be denoted as $s=1-d$. 

Suppose the original study achieved statistical significance at one-sided
level $\alpha$, so $p_o \leq \alpha$.  The standard approach to assess 
replication success is based on significance of the replication effect 
estimate at the same level $\alpha$, \ie the replication is considered successful if also $p_r \leq \alpha$. 
This approach is known in drug development as the two-trials rule \citep{senn:2007}, usually
conducted at $\alpha=0.025$.
Let $z_\alpha = \Phi^{-1}(1-{\alpha})>0$ denote the $z$-value corresponding
to the level $\alpha$, then significance of the replication
\hl{study is achieved if $z_r \geq z_\alpha$, which}
is equivalent to the \hl{condition} 
\begin{equation}\label{eq:dSig}
   d \geq \frac{z_\alpha}{z_o  \, \sqrt{c}}. 
\end{equation}
on the relative effect size \eqref{eq:d}.
The right hand-side goes to zero for increasing $c$, so if the
relative sample size $c$ is large enough, significance of the
replication study can be achieved with any arbitrarily small (but
positive) relative effect size $d$. \hl{However, declaring replication
success when there is substantial shrinkage is contrary to common
sense, as the replication effect estimate may not reflect an effect
size of the same practical relevance as the original one, despite its
statistical significance.}

\hl{In this paper we first review the \citet{held2020} approach for
  the assessment of replication success, followed by showing how it
  relates to the relative effect size (Section \ref{sec:res}). This perspective
  is used in Section \ref{sec:goldenthresh} and \ref{sec:recalib} 
  to propose  a recalibration of the
  method, the \textit{golden level}, which leads to a more appropriate
  criterion for replication success compared to the two-trials rule
  (Section \ref{sec:2TR}). In Section \ref{sec:ER} we study
  power and Type-I error rates of the proposed method and compare it to the
  two-trials rule. The recalibrated method ensures that conditional power can 
  take any desired value if the original study has been significant and the replication
  sample size is large enough
(Section \ref{sec:powerrep}), controls the overall Type-I error if the replication
sample size is not smaller than the original one (Section \ref{sec:T1E}), and 
offers uniform
gains in project power compared to the two-trials rule (Section \ref{sec:PP}). Section \ref{sec:application} 
describes an application to data from four replication projects and Section 
\ref{sec:discussion} closes with some discussion.}

\clearpage
\section{Replication success}\label{sec:RS}

\begin{center}
\begin{figure}[!h]
\begin{center}
\begin{knitrout}
\definecolor{shadecolor}{rgb}{0.969, 0.969, 0.969}\color{fgcolor}

{\centering \includegraphics[width=\maxwidth]{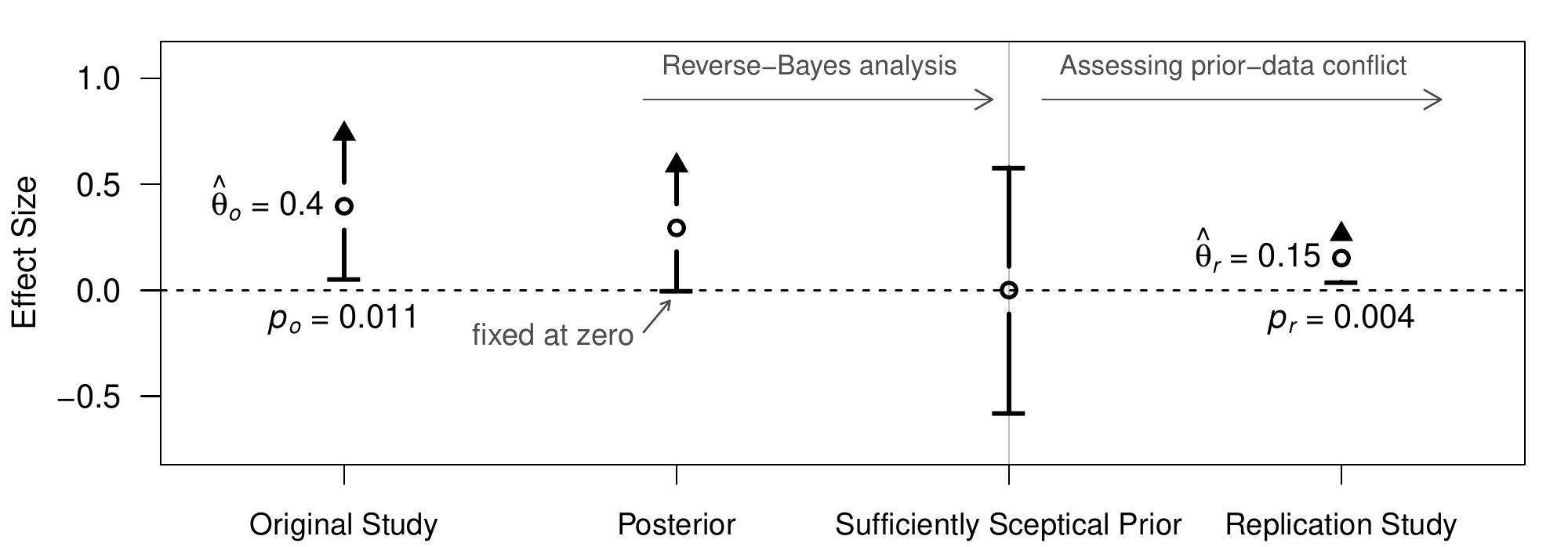} 

}

\end{knitrout}
\caption{Example of the assessment of replication success. The original study from \citet{Pyc2010} has effect estimate  $\hat \theta_o=
  0.4$ on Fisher's $z$ scale
  (95\% CI from $0.05$ to $0.74$) and one-sided $p$-value 
  $p_o = 0.011$.  
  The left part of the figure illustrates the \hl{reverse-Bayes derivation of the sufficiently sceptical prior
    based on}
  the original study result and the posterior with lower credible limit fixed at zero. \hl{The comparison} of the sufficiently sceptical prior with the replication study result ($\hat \theta_r = 0.15$, 95\% CI from 0.04 to 0.26, $p_r = 0.004$) 
  in the right part of the figure \hl{is used to assess potential prior-data conflict.}
  \label{fig:fig1}}
\end{center}
\end{figure}
\end{center}

Hereinafter we focus on the one-sided assessment of replication
success to ensure that replication success can only occur if the
original and replication effect estimates go in the same
direction. Figure \ref{fig:fig1} illustrates the \citet{held2020} approach
based on a replication study
from the \textit{Social Sciences Replication Project} \citep{Camerer2018}: 
the significant original finding by \citet{Pyc2010} at 
one-sided level $\alpha=0.025$ is
challenged with a sceptical prior, sufficiently concentrated around
zero to make the original study result no longer convincing \citep{matthews:2001,matthews:2001b}.
Replication success is then defined as conflict between the
sceptical prior and the result from the replication study in order to
\hl{disprove} the sceptic. Conflict is quantified by a prior-predictive
tail probability $p_{\mbox{\scriptsize Box}}$ \citep{box:1980} where a small value
$p_{\mbox{\scriptsize Box}} \leq \alpha$ defines replication success.
In Figure \ref{fig:fig1}  the original finding is only borderline significant, so the
sufficiently sceptical prior is fairly wide.  Furthermore, there is
substantial shrinkage 
($d = 0.15/0.4 = 
0.38$) 
of the replication effect estimate
and therefore hardly any conflict with the sufficiently
sceptical prior (one-sided $p_{\mbox{\scriptsize Box}}=0.31$).
We are thus not able to declare replication success at
level $2.5$\%.

\hl{The actual value of $p_{\mbox{\scriptsize
      Box}}$ is difficult to interpret as it depends on the level $\alpha$ and does not even exist if
  the original $p$-value $p_o$ exceeds $\alpha$. However, } 
\citet{held2020} showed
that if both $\sign(z_o)=\sign(z_r)$ and
\begin{equation}\label{eq:extrinsic.p}
\left({z_o^2}/{z_{{\alpha_S}}^2}-1 \right) \left({z_r^2}/{z_{{\alpha_S}}^2} 
- 1 \right) \geq c 
\end{equation}
hold, replication success at level ${\alpha_S}$ is achieved, where $z_{{\alpha_S}} = \Phi^{-1}(1-{\alpha_S})$.
The requirement \eqref{eq:extrinsic.p} can be assessed for \hl{any} value of 
\hl{${{\alpha_S}} > \max\{p_o, p_r\}$} and of particular interest is the 
smallest possible value of $\alpha_S$ where \eqref{eq:extrinsic.p} holds, 
the so-called {\it sceptical $p$-value} $p_S$. 
We are thus interested in the value $z_S^2$ that fulfills
\begin{equation}\label{eq:extrinsic.p2}
\left({z_o^2}/{z_{S}^2}-1 \right) \left({z_r^2}/{z_{S}^2} 
- 1 \right) = c. 
\end{equation}
There is a unique solution of \eqref{eq:extrinsic.p2} which defines
the one-sided {sceptical $p$-value} $p_S=
1-\Phi\left({z_S}\right)$ where $z_{S} \coloneqq +
\sqrt{z_{S}^2}$, provided
$\sign(z_o)=\sign(z_r)$ holds.  Replication success at level
$\alpha_S$ is then achieved if $p_S \leq
\alpha_S$.  In the introductory example based on the original study by
\citet{Pyc2010}, the sceptical
$p$-value turns out to be $p_S=0.11$.

The sceptical
$p$-value has a number of interesting properties, see \citet[Section
3.1]{held2020} for details. In particular, $p_S > \max\{p_o, p_r\}$
always holds with $p_S \downarrow \max\{p_o, p_r\}$ for $c \downarrow
0$. Furthermore, if the $p$-values $p_o$ and
$p_r$ are fixed, the sceptical $p$-value
$p_S$ increases with \hl{decreasing} relative effect size
$d$.  The first property ensures that both the original and the
replication study have to be sufficiently convincing on their own to
\hl{achieve} replication success.  The second property guarantees that
shrinkage of the replication effect estimate is penalized.
                                      
The level for replication success $\alpha_S$ has to be distinguished
from the significance level $\alpha$ associated with the ordinary
$p$-value.  \citet{held2020} has used the {\it nominal level} for
replication success ($\alpha_S=\alpha$) for convenience, but in the
following we will propose a recalibration of the procedure along with
a new value for $\alpha_S$, the {\it golden level} (Section~\ref{sec:goldenthresh}). 
\hl{The derivation is based on a property of the required 
relative effect size for replication success, if the relative sample size is very large
(Section~\ref{sec:res}).}
In a nutshell, the golden level
ensures that for original studies which were only borderline
significant \hl{($p_o=\alpha$)}, replication success is only possible if the replication
effect estimate is larger than the original one \hl{($d > 1$)}.\\

\subsection{Relative effect size}\label{sec:res}
Without loss of generality we \hl{now} assume that $\hat \theta_o > 0$ and that
$p_o < {\alpha_S}$ has been observed in the original study, otherwise
it would be impossible to achieve replication success at level
$\alpha_S$ because $p_S$ is always larger than $p_o$. 
The condition \eqref{eq:extrinsic.p} for replication
success can then be re-written as
\begin{equation}\label{eq:eq.z}
  z_r \geq   z_{{\alpha_S}} \sqrt{1+c/(K-1)} \eqqcolon  \zrmin,
\end{equation}
where $K=z_o^2/z_{{\alpha_S}}^2>1$. The right hand-side of
\eqref{eq:eq.z} is the minimum replication $z$-value $\zrmin$ required
to achieve replication success.  Note that $\zrmin$ increases with increasing $c$, so increasing the replication 
sample size leads to a more stringent success requirement \hl{for $z_r$ and the corresponding} replication
$p$-value $p_r$.  

Equation \eqref{eq:eq.z} can be further transformed to a
condition on the relative effect size \eqref{eq:d}:
\begin{equation}\label{eq:res}
  d \geq  \frac{\sqrt{1+c/(K-1)}}{ \sqrt{c K}} \eqqcolon \dmin .
\end{equation}
To achieve replication success, the relative effect size 
must be
at least as large as the right hand-side of \eqref{eq:res}, the
  minimum relative effect size $\dmin$, a function of $K$ and the
relative sample size $c$.  
If the relative sample
size becomes very large, \ie $c \rightarrow \infty$, we have
$\dmin \downarrow \dinfty$ where
\begin{equation}\label{eq:smallestpossible}
  \dinfty = 1/\sqrt{K (K-1)}
\end{equation}
\hl{is the {\it limiting relative effect size}.}
This shows that the minimum relative effect size $\dmin$ in \eqref{eq:res}
does not go to zero for increasing $c$, so replication success cannot
be achieved if the relative effect size $d$ is smaller or equal to
$\dinfty$, no matter how large the replication study is.  
\hl{In contrast}, the corresponding criterion \eqref{eq:dSig} \hl{of the 
two-trials rule}
can be achieved for any positive relative effect size, regardless of
how small, provided the replication sample size is sufficiently large.

\subsection{The golden level}\label{sec:goldenthresh}
Significance of both the original and the replication study at level
$\alpha$ is a necessary but not sufficient requirement for replication
success at the nominal level ($\alpha_S = \alpha$).  The nominal level
may therefore be too stringent.  It is more reasonable to calibrate
the procedure in such a way that to establish replication success,
original and replication study do not both necessarily need to be
significant at level $\alpha$, provided that the replication effect
estimate does not shrink compared to the original one.  We therefore 
choose a level $\alpha_S$ such that a borderline
significant original study ($p_o = \alpha$) cannot lead to replication
success if there is shrinkage $s > 0$ of the replication effect
estimate.  Mathematically, this translates to setting $\dinfty = 1$
and $K = z_\alpha^2/z_{\alpha_S}^2$ in~\eqref{eq:smallestpossible} and
leads to the quadratic equation $K(K-1) = 1$ with solution
$K = \varphi$ where
$\varphi = (\sqrt{5}+1)/2 \approx 1.62$ is
known as the golden ratio. Solving for $z_{\alpha_S}$ gives
\hl{$z_{\alpha_S} = z_\alpha / \sqrt{\varphi}$
and the corresponding  golden level 
\begin{eqnarray}\label{eq:golden}
\alpha_S & = & 1 -  \Phi(z_\alpha / \sqrt{\varphi} )
\end{eqnarray}
for replication success. }
This is our recommended default choice to assess replication success and we will study its properties in the following
                   in more detail. For $z_\alpha = 1.96$ (one-sided $\alpha = 0.025$), 
the golden level is $\alpha_S =0.062$.
In the introductory example 
shown in Figure \ref{fig:fig1}, 
the sceptical $p$-value is $p_S=0.11 > 0.062$, so 
the replication  of the \citet{Pyc2010} study was not successful.

\hl{The golden level \eqref{eq:golden} is derived from~\eqref{eq:smallestpossible} with $\dinfty = 1$. 
However, we may also use a different value for the 
limiting relative effect size $\dinfty$, say $\dinfty=0.8$. 
Then replication success is only possible for a borderline significant result ($p_o = \alpha$) 
if there is less than $1-\dinfty$ (20\% for $\dinfty=0.8$) shrinkage of the replication effect estimate. 
This approach is equivalent to a limiting relative effect size of 1 if the original $p$-value
$p_o$ is equal to a different level $\alpha'$, which can be derived as follows:  
First, solving \eqref{eq:smallestpossible} for $\dinfty > 0$ gives
$K = z_\alpha^2/z_{\alpha_S}^2 = 1/2+\sqrt{1/4 + 1/\dinfty^2}$. The new level
$\alpha'$ fulfills  $\varphi=z_{\alpha'}^2/z_{\alpha_S}^2$, so $z_\alpha^2/K=z_{\alpha'}^2/\varphi$
and therefore
\begin{equation}\label{eq:alphaPrime}
\alpha' = 1-\Phi\left(z_{\alpha} \, \sqrt{\varphi / K} \right).
\end{equation}
For example, for $\alpha = 0.025$ and $\dinfty = 0.8$ we obtain $\alpha'=0.033$. 
}
   
\subsection{Recalibration of the sceptical $p$-value}\label{sec:recalib}

The condition $p_S \leq \alpha_S$ for replication success at the
golden level is equivalent to $z_S \geq z_\alpha / \sqrt{\varphi}$,
\ie $z_S \sqrt{\varphi} \geq z_\alpha$.  In practice it may be
preferable to recalibrate the sceptical $p$-value
$p_{S} = 1 - \Phi(z_S)$ to
$ \tilde{p}_S = 1 - \Phi(z_S \sqrt{\varphi})$, which then needs to be
compared to $\alpha$ (rather than $\alpha_S$) to assess replication
success and can thus be interpreted on the same scale as an ordinary
$p$-value.  For example, the recalibrated sceptical $p$-value for the
replication of \citet{Pyc2010} turns out to be
$\tilde{p}_S=0.061$ and does not lead to replication
success at \hl{any level $\alpha<0.061$, including} the
standard $0.025$ level.

\subsection{Comparison with the two-trials rule}\label{sec:2TR}
A useful benchmark for comparison is the two-trials rule in drug
development \citep[Section 9.4]{kay:2015}, which requires ``at least
two adequate and well-controlled studies, each convincing on its own,
to establish effectiveness'' \citep[p.~3]{FDA1998}.  This is usually
achieved by independently replicating the result of a first study in a
second study, both significant at one-sided level
$\alpha=0.025$.  {It is worth noting that in practice the two trials are often run in parallel} {\citep{senn:2007}}{, so do not exactly resemble the replication setting.}

\begin{figure}[!h]
\centering
\begin{knitrout}
\definecolor{shadecolor}{rgb}{0.969, 0.969, 0.969}\color{fgcolor}

{\centering \includegraphics[width=\maxwidth]{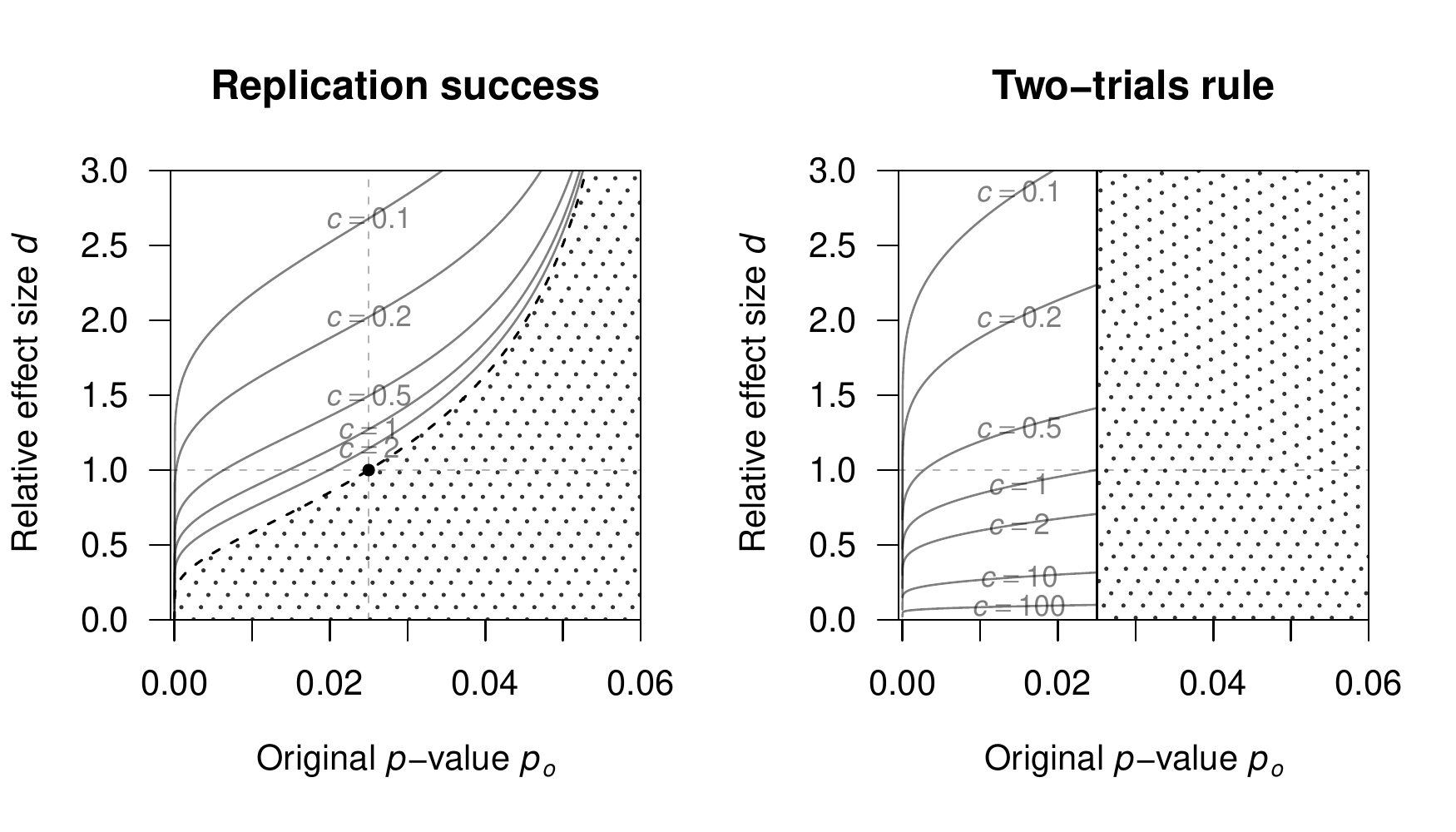} 

}

\end{knitrout}
\caption{Comparison of replication success at the golden level ($p_S \leq \alpha_S = 
0.062$) and the two-trials rule ($p_o \leq 0.025$ and $p_r \leq 0.025$). 
The \hl{dotted} areas indicate that success is
impossible for original $p$-value $p_o$ and relative effect size $d$. 
In the white areas success is possible and
depends on the relative sample size $c$ as indicated by the grey lines. 
\hl{The dashed black line in the left plot indicates the limiting relative 
effect size $\dinfty$.}
}
\label{fig:fig2}
\end{figure}

The main difference between the replication success and the two-trials
rule approach concerns how shrinkage of the replication effect
estimate is handled. Figure~\ref{fig:fig2} illustrates that
shrinkage is penalized in the assessment of replication success, \ie
the original $p$-value needs to be quite small to achieve replication
success for a relative effect size $d<1$. In contrast, significance of
the replication study can be achieved even if there is substantial
shrinkage, provided the replication sample size is large enough.

It is interesting to directly compare the two-trials rule and
replication success at the golden level in terms of 
the required relative effect size $d$ to fulfill the criteria
\eqref{eq:dSig} and \eqref{eq:res}, respectively, see Figure~\ref{fig:fig2}.  If the original
$p$-value is not significant at level $\alpha$, only replication
success can be achieved, but will require a replication effect
estimate larger than the original one. 
\hl{
  For example, four studies with one-sided
  $p_o \in (0.025, 0.03)$ have been included in the
  \textit{Reproducibility Project: Psychology} \citep{OSC2015} and one
  of them achieves replication success (see Section \ref{sec:application} for
  details).  By definition, such non-significant original findings can
  never fulfill the two-trials rule.
}

If the original $p$-value is
smaller than $\alpha$, then the situation depends on the relative
sample size $c$.  For example, when the replication sample size is
chosen to be the same as in the original study ($c=1$) and
$\alpha=0.025$, original studies with a $p$-value larger than
$0.006$ will require a smaller relative effect
size $d$ with the two-trials rule, while $p$-values smaller than
$0.006$ will require a smaller relative effect
size $d$ with the replication success method.  This illustrates that
the latter method is less stringent than the two-trials rule if the
original study is already sufficiently convincing.

\section{Power and Type-I Error Rate}\label{sec:ER} 
Although Bayesian methods do not rely on the frequentist paradigm of
repeated testing, it is still useful to investigate their frequentist
operating characteristics \citep{Dawid1982, Rubin1984, Grieve2016}
and this also holds for the proposed reverse-Bayes assessment of
replication success. We first condition on the results from the original
study and compare the power to achieve replication success with the two-trials
rule in Section \ref{sec:powerrep}. 
We then
assume that none of the two studies have been conducted and
investigate the \hl{overall} Type-I error rate (Section \ref{sec:T1E}) and the project power
(Section \ref{sec:PP}) 
\citep{Maca2002} \hl{over both studies in combination for fixed relative sample size $c$}.

\subsection{Conditional power}\label{sec:powerrep}

\begin{figure}[!ht]
\begin{center}
\begin{knitrout}
\definecolor{shadecolor}{rgb}{0.969, 0.969, 0.969}\color{fgcolor}

{\centering \includegraphics[width=\maxwidth]{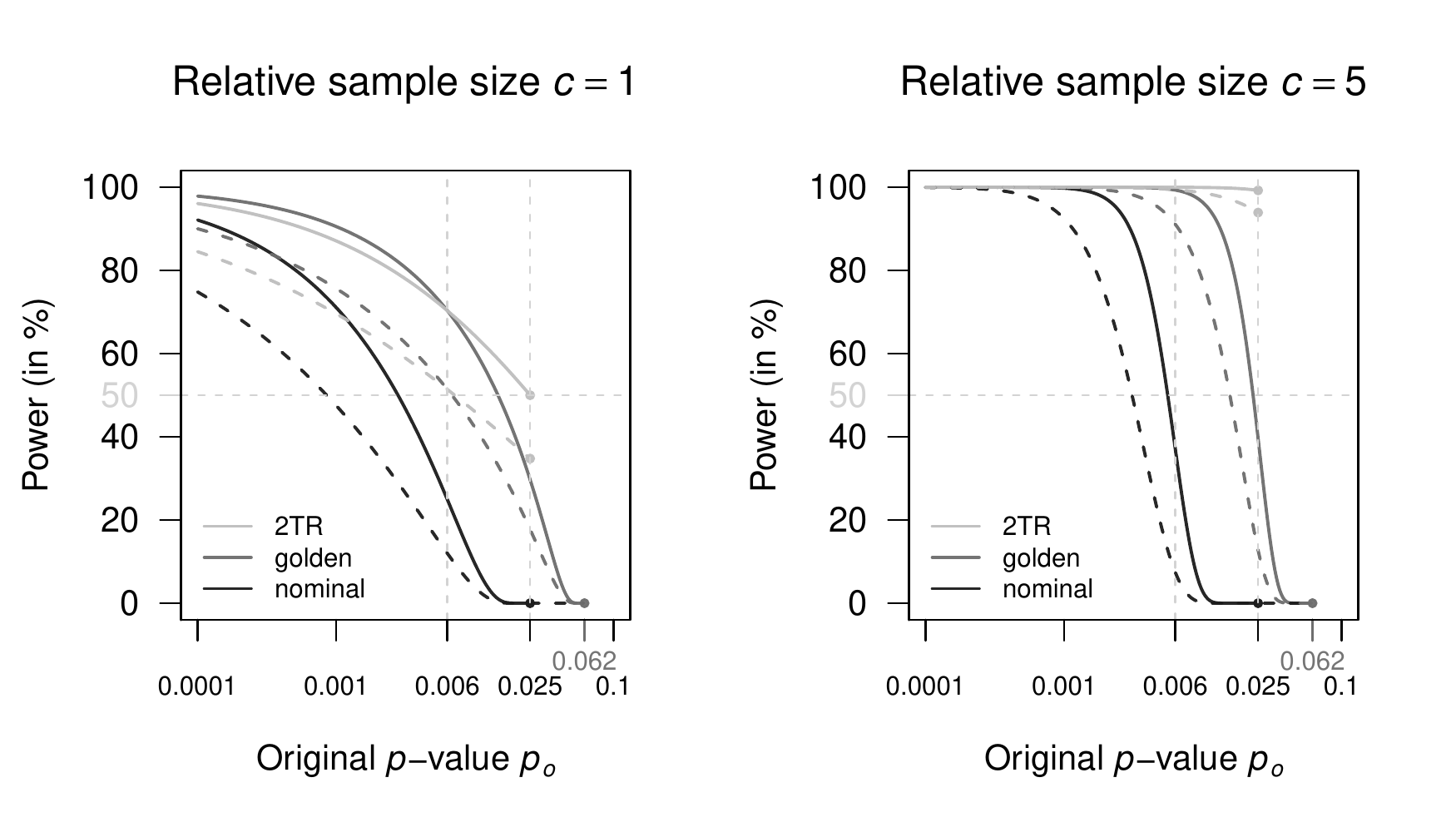} 

}

\end{knitrout}
\end{center}
\caption{\hl{Conditional} power as a function of the {one-sided}
  $p$-value of the original study with \hl{relative} sample size
  $c=1$ (left) and $c=5$ (right). Shown is
  conditional power \hl{assuming the unknown parameter is equal to 
  the original effect estimate (solid)} \hl{and conditional
    power based on 20\% shrinkage of the original
    effect estimate (dashed)} for the two-trials rule (2TR) at level
  $\alpha = 0.025$ and for replication success 
  at the
  corresponding golden and nominal level. Power values of exactly zero
  are omitted.}
\label{fig:fig3}
\end{figure}

Figure~\ref{fig:fig3} compares the power for replication success
\hl{\citep[see][Section 4 for details]{held2020}} at the golden and at
the nominal level with the power of the two-trials rule for relative
sample size $c = 1$ (left) and $c = 5$ (right)
as a function of the one-sided $p$-value \hl{$p_o$} from the original
study.  \hl{Shown is the conditional power assuming the unknown parameter
  $\theta$ is equal to the original effect estimate $\hat \theta_o$. Then
$\hat \theta_r \given \hat \theta_o \sim \Nor(\hat \theta_o, \kappa^2/n_r)$
and it follows that
$d \given \hat \theta_o \sim \Nor(1, 1/(c z_o^2))$. 
The conditional power for replication
success can therefore be calculated as 
\begin{equation}\label{eq:condPower}
\Pr(d \geq \dmin \given \hat \theta_o) = \Phi\left[\sqrt{c} z_o (1-\dmin)\right]
\end{equation}
where $\dmin$ is
given in \eqref{eq:res}.  Predictive power, which is conditional
power averaged over a $\Nor(\hat \theta_o, \sigma_o^2)$ distribution
for the effect size $\theta$, could also be calculated, then
\mbox{$d \given \hat \theta_o \sim \Nor(1, (1+1/c)/z_o^2)$}. 
Conditional and predictive power of the two-trials rule 
also depend on $z_o$, $c$ and $\alpha$ and are given in
\citet{MicheloudHeld2020}.}

The two-trials rule requires a significant original study and hence it is 
impossible to power a replication study when $p_o > 0.025$. 
\hl{The same applies for replication success at the nominal level,
where the power is zero for any $p_o > 0.025$, regardless
of the replication sample size.
This is different for the golden level,
where the conditional power of an original study with $0.025 < p_o 
< 0.062$ is low, but not zero.}
\hl{However, if the original $p$-value $p_o$ is slightly smaller than
$0.025$, the two-trials rule has a larger power, both for $c=1$ and $c=5$. }
\hl{But if the original $p$-value is sufficiently
small ($p_o < 0.006$ for $c=1$), the power for replication success at the golden level is larger
than the power of the two-trials rule. }

\hl{ Compared to
  $c=1$, the conditional power for $c=5$
  of both the two-trials rule 
  and the replication success approach at the golden level 
  increases if $p_o \leq \alpha$.     A remarkable feature of the
  replication success approach at the golden level is that conditional
  power can be pushed towards 100\% for large enough $c$ if
  $p_o<\alpha$, but not otherwise. This can be seen from \eqref{eq:condPower} because 
  $\dmin < 1$ for $p_o<\alpha$ and large enough relative sample size $c$.
On the other hand, for $p_o > \alpha$ conditional power for replication success
will tend to 0\% for increasing $c$ because $\dmin>1$ for all $c$. Finally, for $p_o = \alpha$
the limit is 50\%. 
  The same
  property can be observed at the nominal level, however at the
  smaller threshold $1-\Phi(z_\alpha \sqrt{\varphi})$ which is $0.006$ for  $\alpha=0.025$. Only if $p_o < 0.006$ will the conditional power 
  for replication success attain 100\% for $c \to \infty$. This further highlights the stringency
  of the nominal level.  }

\hl{The approach described so far takes the original study at
  face-value since it assumes that $\hat \theta_o$ is equal to the unknown effect size $\theta$. 
  In practice, however,
  there are often good reasons to believe that original effect estimates
  have a tendency to be inflated (\eg due to publication bias). One
  way to address this issue is to base power calculations on a
  shrunken version of the original effect
  estimate, where the amount of shrinkage is guided by domain knowledge and 
  a risk of bias assessment of the original study.
  For illustration, Figure~\ref{fig:fig3} also shows conditional power based
  on 20\% shrinkage of the original effect estimate
  which reduces the conditional power for all methods, especially for
  a relative sample size $c = 1$.  Conditional power for
  replication success at the golden level can now be pushed towards 100\%
  only for $p_o < 0.018$, which can be derived by
  solving \eqref{eq:alphaPrime} for $\alpha$ with
  $\alpha'=0.025$ and $\dinfty=0.8$.  To be able to
  push conditional power based on 20\% shrinkage towards 100\% for all
  $p_o < 0.025$, 
  equation \eqref{eq:alphaPrime} would have to be used directly to relax the level
  from $\alpha=0.025$ to $\alpha'=0.033$. 
  }

\subsection{Overall Type-I error rate}\label{sec:T1E}
\hl{The two studies are assumed to be independent with Type-I error
rate fixed at $\alpha$ for each of them,} so the Type-I error rate of the
two-trials rule \hl{over the entire project} is simply $\alpha^2$ for
any value of the relative effect size $c$.
In contrast, the Type-I
error rate of the proposed replication success assessment {depends on}
\hl{the relative sample size} $c$.

For $c=1$, \citet[Section 3]{held2020} showed that $z_S^2$ in \eqref{eq:extrinsic.p2}
simplifies to half the harmonic mean of the squared test statistics
$z_o^2$ and $z_r^2$.  The
connection $z_S^2 = z_H^2/4$ to the harmonic mean $\chi^2$-test
statistic $z_H^2$ \citep{held2020b}, which has a
$\chi^2(1)$-distribution under the null hypothesis, makes it
straightforward to compute the Type-I error rate  at level
$\alpha_S$ for $c = 1$ as 
\begin{eqnarray}\label{eq:T1E}
\mbox{T1E} =   \left\{1-\Phi\left[2 \, \Phi^{-1}\left(1-\alpha_S \right)
 \right]\right\}/2.
\end{eqnarray}
For the golden level $\alpha_S =0.062$ at
$\alpha = 0.025$, the Type-I error rate \eqref{eq:T1E} 
is $0.0515$\%, slightly less than the Type-I
error rate $\alpha^2=0.0625$\% of the
two-trials rule.  For comparison, the Type-I error rate at the nominal
level $\alpha_S=0.025$ is
$0.0022$\%, \hl{much} smaller than
0.0625\%.

For $c \neq 1$,  the Type-I error rate can be calculated through numerical 
integration: 
\begin{equation}\label{eq:T1E0}
  \mbox{T1E} = \int_{z_{\alpha_S}}^{\infty}  
\P(z_r \geq \zrmin \given z_o, c, \alpha_S) \, 
  \phi(z_o) \, dz_o,
\end{equation}
where $\phi(\cdot)$ denotes the standard normal density function. 
The first term in the integral of \eqref{eq:T1E0} is the probability of 
replication success at level $\alpha_S$ conditional on a fixed original test 
statistic $z_o$ and a relative sample size $c$. 
Now $z_r \sim \Nor(0,1)$ under the null hypothesis, so 
this term simplifies to 
  $\P(z_r \geq \zrmin \given z_o, c, \alpha_S) = 1 - \Phi(\zrmin)$
where $\zrmin$ in \eqref{eq:eq.z} depends on $z_o$, $c$, and $\alpha_S$.

\begin{figure}[!h]
\centering

\begin{knitrout}
\definecolor{shadecolor}{rgb}{0.969, 0.969, 0.969}\color{fgcolor}

{\centering \includegraphics[width=\maxwidth]{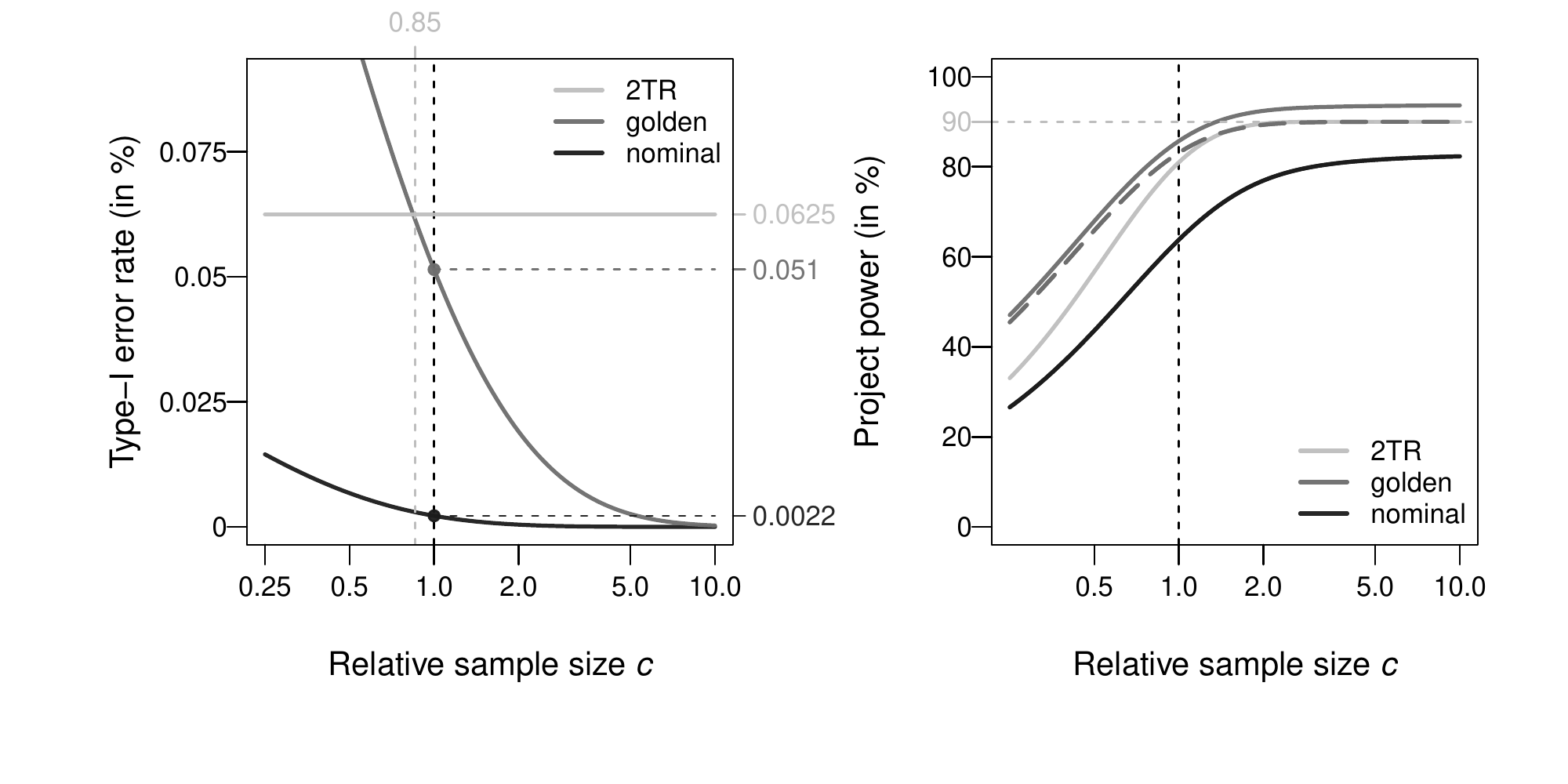} 

}

\end{knitrout}

\caption{\hl{Overall} Type-I error rate (left) and project power (right) for fixed relative sample size $c$. 
  Results are given for replication success 
  at the nominal and golden
  level and compared with the two-trials rule (2TR) at $\alpha = 0.025$.
  \hl{The dashed darkgrey line is the project power at the golden level based on significant original studies ($p_o \leq 0.025$).}
  The power of the original study is 90\%}
\label{fig:fig4}
\end{figure}

The left plot in Figure \ref{fig:fig4} displays the Type-I error
rate for $\alpha=0.025$ as a function of the relative sample
size $c$.  It can be seen that the Type-I error \hl{of the replication success approach} decreases with
increasing relative sample size $c$. \hl{This also follows from \eqref{eq:T1E0} 
  where
$\P(z_r \geq \zrmin \given z_o, c, \alpha_S) = 1 - \Phi(\zrmin)$ 
decreases with increasing $c$, because $\zrmin$ increases with increasing $c$,
see equation \eqref{eq:eq.z}. }

The Type-I error rate of the nominal level is always below the target
$0.0625$\%.  Although the Type-I error will
eventually attain $\alpha^2$ in the limit $c \downarrow 0$
\citep[Section 3.4]{held2020}, the nominal level seems to be too
stringent for realistic values of $c$.  The Type-I error rate of the
golden level is smaller than $0.0625$\% for
$c >0.85$.  Appropriate Type-I error control is
thus ensured even for replication studies where the sample size is
slightly smaller than in the original study. 
  
\begin{figure}[!ht]
\begin{center}

\begin{knitrout}
\definecolor{shadecolor}{rgb}{0.969, 0.969, 0.969}\color{fgcolor}

{\centering \includegraphics[width=\maxwidth]{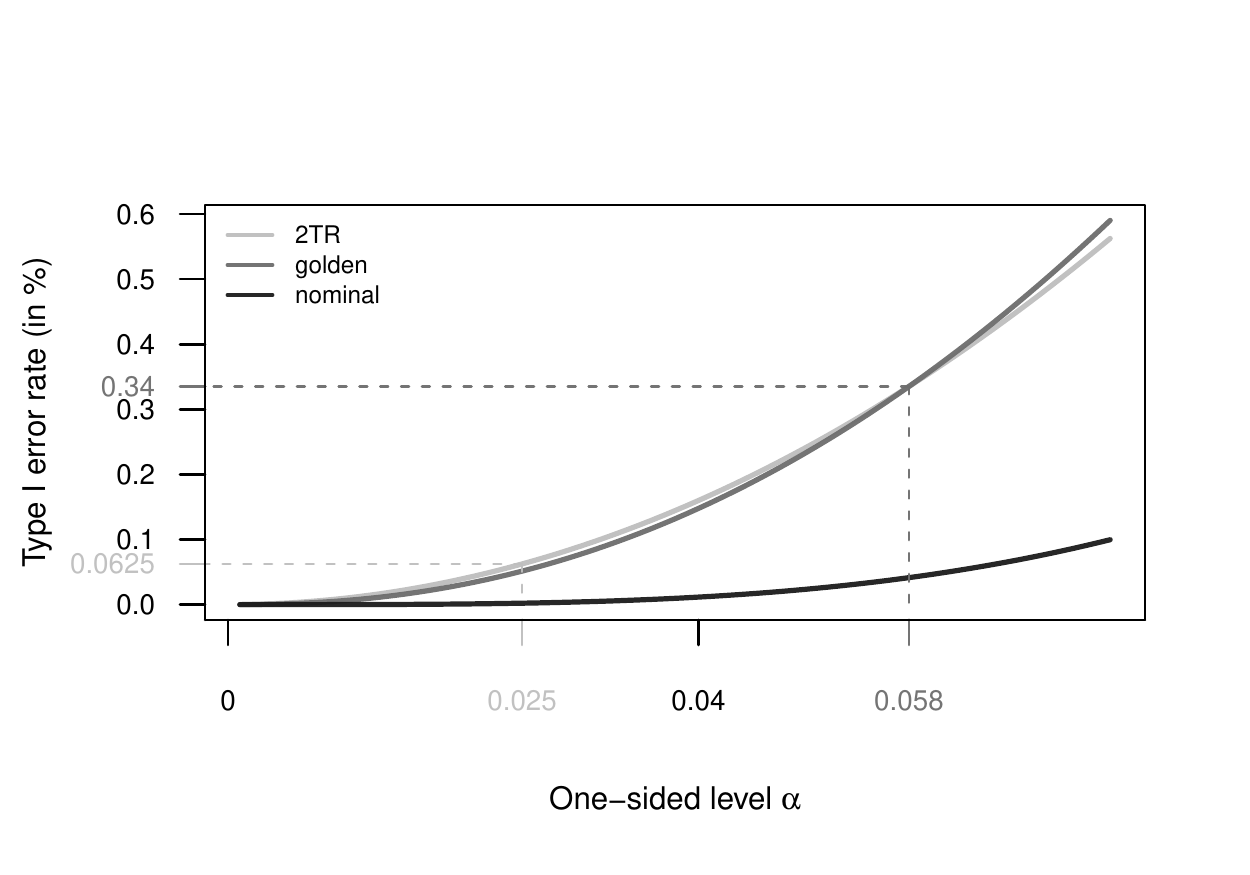} 

}

\end{knitrout}

\end{center}
\caption{\hl{Overall Type-I error rate if the
  replication sample size equal to the original study
  ($c=1$).  The two-trials rule (2TR) 
  is compared to replication success 
  at the
  golden and nominal level for different values of $\alpha$.}}
\label{fig:fig5}
\end{figure}

\hl{Figure \ref{fig:fig5} compares for $c=1$ the Type-I error rate \eqref{eq:T1E}
  of replication success at the golden and at the nominal level 
    with the
  two-trials rule for different values of $\alpha$. The
  Type-I error rate of the two-trials rule is $\alpha^2$ and the
  replication success approach at the nominal level always has a much
  smaller Type-I error rate than $\alpha^2$.  At the golden level
  the Type-I error rate of the 
  replication success approach is much closer to $\alpha^2$, still slightly smaller 
  if $\alpha < 0.058$. For $\alpha = 0.058$ the
  Type-I error rate is equal to the
  Type-I error rate $0.058^2 = 0.34\%$ of the 
  two-trials rule and for $\alpha > 0.058$ the 
  Type-I error rate is slightly larger than $\alpha^2$.
  The
  Type-I error rate for replication success decreases with
  increasing $c$, so as long as the replication sample size is not
  smaller than the original sample size, Type-I error control at
  $\alpha^2$ is guaranteed at the golden level for any one-sided level
  $\alpha < 0.058$.} 

\subsection{Project power}\label{sec:PP}
Under the alternative we have
$z_o\sim \Nor(\mu, 1)$ with $\mu = z_\alpha + z_\beta$
where $\alpha$ is
the assumed significance level and 
\hl{
$1-\beta = \Phi(\mu - z_\alpha)$ 
is the power to detect the
assumed effect $\theta = \mu \sigma_o$ in the original study \citep[Section 3.3]{mat2006}.  
In the following
$\alpha=0.025$ and $\beta=0.1$ are
used. The power of a significant
replication study with sample size $n_r = c n_o$ is 
\begin{equation*}
\Phi(\theta/\sigma_r - z_\alpha) = 
\Phi(\sqrt{c} \mu - z_\alpha),
\end{equation*}
}
so 
depends on $\mu$ and the relative sample size $c$. The project power of the
two-trials rule is therefore $(1-\beta) \, \Phi(\sqrt{c} \mu - z_\alpha)$ \hl{and increases with increasing $c$.}

The project power for replication success is computed as 
\begin{equation*}
  \mbox{PP} = \int_{z_{\alpha_S}}^{\infty}  
\P(z_r \geq \zrmin \given z_o, c, \alpha_S) \, 
  \phi(z_o - \mu) \, dz_o
\end{equation*}
and shown in the right plot of Figure \ref{fig:fig4} as a function
of $c$.  For the golden level, the project power quickly increases to
values above 90\%, whereas the nominal level only reaches around 80\%
project power. The project power based on the two-trials rule is shown
for comparison, which is always smaller than for the golden level 
and converges to 90\% for large $c$.  

\hl{The advantage in power stems partly from replication success still
  being possible when the original $p$-value is larger than 0.025, but
  smaller than 0.062. If we assume that a replication study is
  only conducted if the original study is significant (with
  $p_o \leq 0.025$), then the project power based on the
  golden level (the dashed line in Figure \ref{fig:fig4}) is
  slightly smaller and for $c>1$ barely different than for the
  two-trials rule.  More substantial gains are still visible for
  $c < 1$. 
  However, the restriction to original studies with
  $p_o \leq 0.025$ may not reflect current practice in
  large-scale replication projects. For example, 5 out of
  143 replication studies considered in Section
  \ref{sec:application} do have original $p$-values between
  0.025 and 0.062.  }

\section{Application}\label{sec:application}
In this section, we illustrate the proposed methodology using 
data from four replication projects.
All four projects reported effect estimates that were transformed to
correlation coefficients ($r$). This scale allows for easy comparison of 
effect estimates from studies that investigate different phenomena
\hl{and} is bounded to the interval between minus one and one.
Moreover, the Fisher $z$-transformation $\hat{\theta} = \text{tanh}^{-1}(r)$
can be applied to the correlation coefficients, resulting in the transformed 
estimates being asymptotically normal with variance which is only a function
of the study sample size $n$, \ie $\Var(\hat{\theta}) = 1/(n - 3)$
\citep{Fisher1921}.

The first data set comprises the results from the \textit{Reproducibility Project:
Psychology} \citep{OSC2015}, whose aim was to replicate 100 studies, all of which 
were published in three major Psychology journals in 2008. For our purpose only 
the 73 study pairs from the ``meta-analytic'' subset are considered, since only 
for these studies the standard error of the Fisher $z$-transformed effect 
estimates can be computed \citep{Johnson2016}. 
The second data set comes from the \textit{Experimental Economics Replication 
Project} \citep{Camerer2016} which attempted to replicate 18 experimental 
economics studies published in two high impact economics journals between 2011 
and 2015.
The third data stem from the \textit{Social Sciences Replication Project} 
\citep{Camerer2018} where 21 replications of studies on the social sciences 
were carried out, all of which were originally published in the journals
\textit{Nature} and \textit{Science} between 2010 and 2015.
The last data set originates from the \textit{Experimental Philosophy Replicability
Project} \citep{Cova2018} which involved 40 replications of studies from the
emerging field of experimental philosophy. Since only for 31 studies effective 
sample size for original and replication study were available simultaneously, 
only these pairs were included. For more information on the data sets see
also \citet{Pawel2020}.

Table~\ref{tab:marginalres} presents overall results for each of the 
replication projects. While the median relative effect size is below one for 
all of the four projects, there are still large differences. For
example, the median relative effect size is only 
$0.29$
in the Psychology project, whereas it is 
$0.86$ in the Philosophy project. The degree of shrinkage is also 
reflected in the success rates (according to the two-trials rule and the replication
success approach \hl{at the golden level}), which are around 30\% for the former and more than 70\%
for the latter. The proportion of successful replications is similar for the 
two-trials rule and the replication success approach. \hl{In the Experimental 
Economics project the methods perfectly agree, while in the other three projects
the methods disagree for a few studies.}

\begin{table}[!ht]
\caption{Results for each replication project: 
Relative effect size $d$ (median with 25\% and 75\% quantiles on Fisher's $z$ scale), proportion of 
successful replications with the two-trials rule (2TR) and the replication success (RS)
approach (at the golden level), and number of studies where the methods disagree.}
\resizebox{\textwidth}{!} {
\begin{tabular}{lcccc}
  \toprule
Project & relative effect size $d$ & 2TR (\%) & RS (\%) & discrepant \\ 
  \midrule
Psychology & 0.29 [0.03, 0.77] & 28.8 & 30.1 & 3/73 \\ 
  Experimental Economics & 0.67 [0.35, 0.92] & 55.6 & 55.6 & 0/18 \\ 
  Social Sciences & 0.52 [0.13, 0.65] & 61.9 & 52.4 & 2/21 \\ 
  Experimental Philosophy & 0.86 [0.47, 1.12] & 74.2 & 71.0 & 1/31 \\ 
   \bottomrule
\end{tabular}

}
\label{tab:marginalres}
\end{table}

\begin{figure}[!h]
\centering
\begin{knitrout}
\definecolor{shadecolor}{rgb}{0.969, 0.969, 0.969}\color{fgcolor}

{\centering \includegraphics[width=\maxwidth]{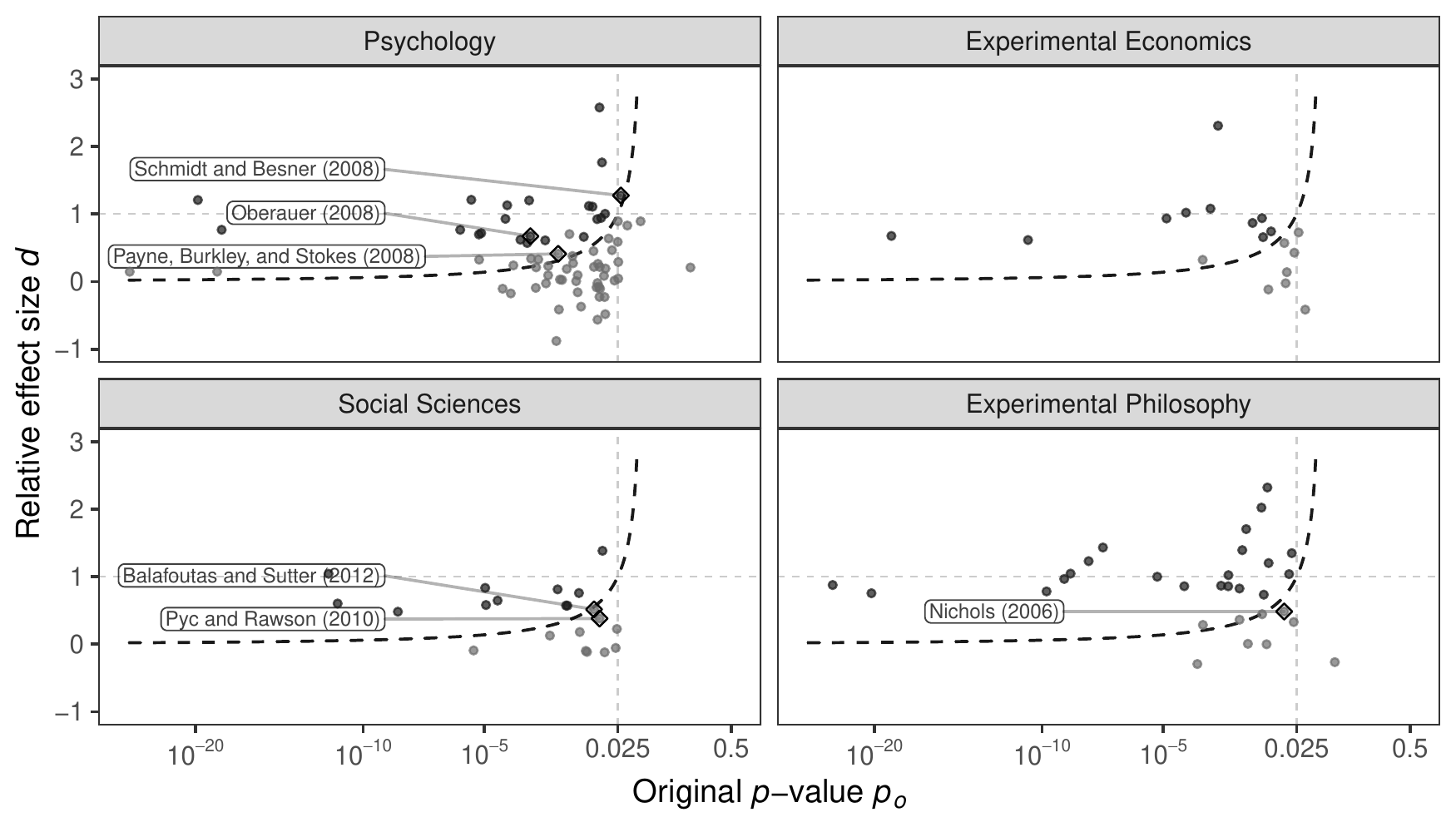} 

}

\end{knitrout}
\caption{Relative effect size $d$ versus original $p$-value $p_o$. 
Black indicates that replication success was achieved at the golden level
while grey indicates that it was not. The diamonds mark studies where 
the replication success approach (at the golden level) and the two-trials rule disagree. The dashed
black line indicates the \hl{limiting relative effect size}
at the golden level with $\alpha = 0.025$. 
}
\label{fig:fig6}
\end{figure}

Figure~\ref{fig:fig6} displays the relative effect size $d$ versus the 
original $p$-value $p_o$ for each study pair and stratified by project.
Note that one study pair from the Philosophy project is not shown due to
extremely small original $p$-value and another study pair from the Psychology project 
is not shown due to a very large relative effect size. 
We can see that for most of the study pairs, the replication success approach 
and the two-trials rule lead to the same conclusion, only six replications 
show conflicting results.
They are highlighted with diamonds in Figure~\ref{fig:fig6} and their
characteristics are summarised in Table~\ref{tbl:discrep}.
\begin{table}[!ht]
\caption{Characteristics of studies for which the replication success approach
(at the golden level)
and the two-trials rule disagree (at one-sided $\alpha = 0.025$). Shown are relative
sample size $c$, relative effect size $d$, original, replication and recalibrated sceptical 
$p$-value $p_o$, $p_r$ and $\tilde{p}_S$.}
\label{tbl:discrep}
\resizebox{\textwidth}{!} {
\begin{tabular}{lllllll}
  \toprule
Study & Project & $c$ & $d$ & $p_o$ & $p_r$ & $\tilde{p}_S$ \\ 
  \midrule
\citet{Schmidt2008} & Psychology & 2.58 & 1.28 & \textbf{0.028} & \textcolor{black}{< 0.0001} & \textcolor{black}{0.024} \\ 
  \citet{Oberauer2008} & Psychology & 0.60 & 0.67 & \textcolor{black}{0.0003} & \textbf{0.035} & \textcolor{black}{0.017} \\ 
  \citet{Payne2008} & Psychology & 2.65 & 0.41 & \textcolor{black}{0.001} & \textcolor{black}{0.023} & \textbf{0.031} \\ 
  \citet{Balafoutas2012} & Social Sciences & 3.48 & 0.52 & \textcolor{black}{0.009} & \textcolor{black}{0.011} & \textbf{0.04} \\ 
  \citet{Pyc2010} & Social Sciences & 9.18 & 0.38 & \textcolor{black}{0.011} & \textcolor{black}{0.004} & \textbf{0.061} \\ 
  \citet{Nichols2006} & Experimental Philosophy & 9.40 & 0.49 & \textcolor{black}{0.015} & \textcolor{black}{0.0006} & \textbf{0.049} \\ 
   \bottomrule
\end{tabular}

}
\end{table}
Two studies from the Psychology project show replication success 
but fail the two-trials rule. 
These studies show $p$-values that are slightly above the significance
threshold in either original or replication study, but do not exhibit much 
shrinkage;
In the replication of \citet{Oberauer2008}, the replication $p$-value was 
$p_r = 0.035$,
a little too large to pass the two-trials rule. However, as the replication 
effect estimate shrunk only about 
$30$\%
compared to the original one, replication success is still achieved. Conversely,
the original $p$-value 
$p_o = 0.028$
in \citet{Schmidt2008} was just above the significance level, yet the 
replication led to a highly significant result
$p_r < 0.0001$
with the effect estimate being even
$30$\%
larger than the original counterpart, which therefore also resulted in 
replication success.

The remaining conflicting studies do not show replication 
success despite passing the two-trials rule. 
In all cases, there is substantial shrinkage of the replication effect estimate
compared to the original one. For instance, in the replication study of 
\citet{Pyc2010}, the estimate shrunk by 
$62$\% 
and the replication $p$-value was only significant because the 
sample size was increased by a factor of 
$c = 9.2$.

\begin{figure}[!htb]
\begin{knitrout}
\definecolor{shadecolor}{rgb}{0.969, 0.969, 0.969}\color{fgcolor}

{\centering \includegraphics[width=\maxwidth]{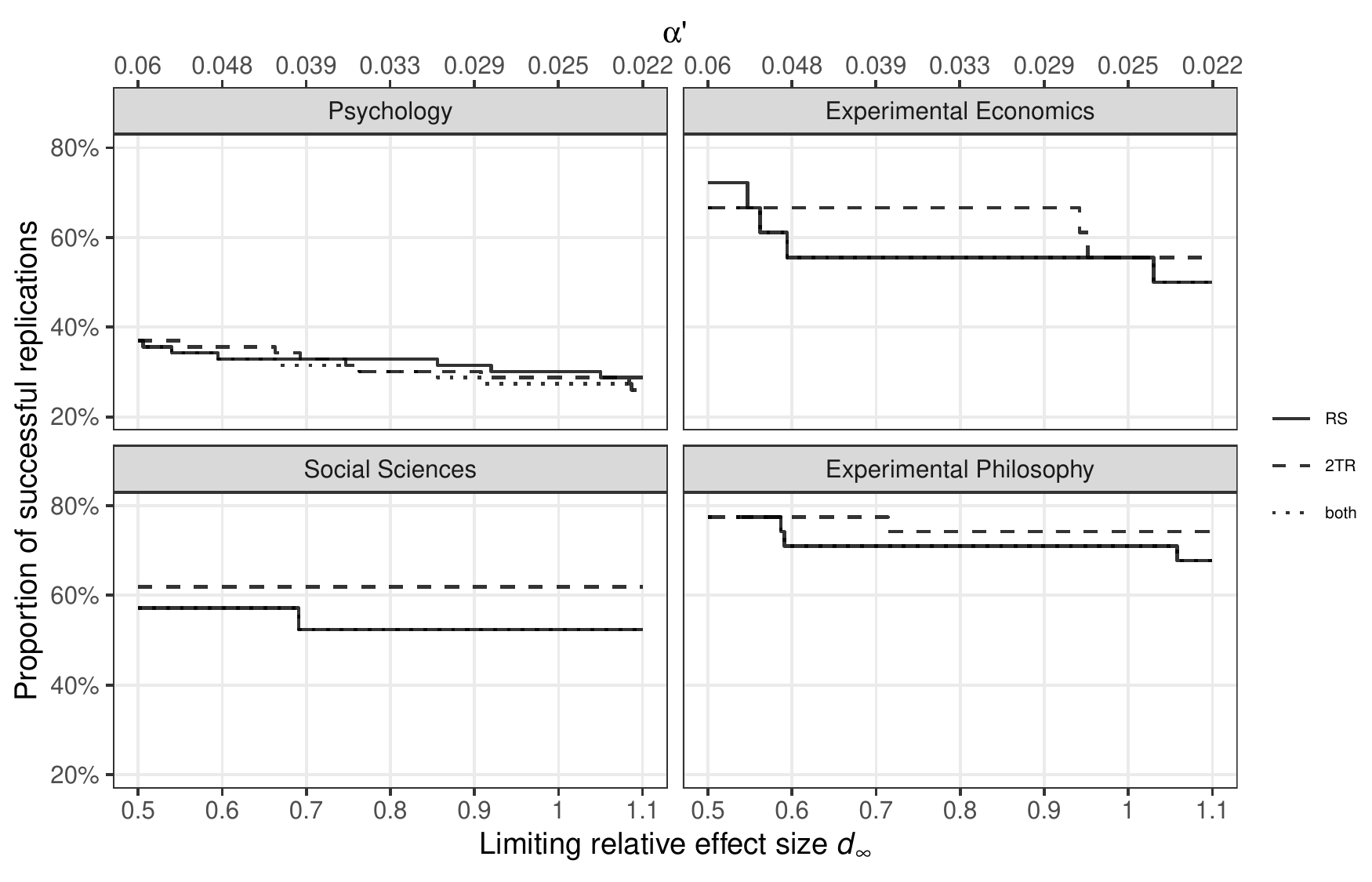} 

}

\end{knitrout}
\caption{Proportion of successful replications as a function of the limiting
relative effect size $\dinfty$ at $\alpha = 0.025$. The upper axis gives the 
equivalent level $\alpha'$ where the corresponding limiting relative effect size is 1. 
The replication success (RS) approach is compared with the two-trials rule (2TR).}
\label{fig:fig7}
\end{figure}

\hl{This analysis was based on the default choice $\dinfty=1$
  at $\alpha=0.025$ for the golden level as described in Section
  \ref{sec:goldenthresh}. We may also choose a different value for the
  limiting relative effect size $\dinfty$ at $\alpha=0.025$ which then
  corresponds to $\dinfty=1$ at a different level $\alpha'$ as given in \eqref{eq:alphaPrime}.  Figure
  \ref{fig:fig7} compares the proportion of successful
  replications with the replication success approach for
  $\dinfty \in (0.5, 1.1)$ with the two-trials rule at the
  corresponding levels
  $\alpha' \in (0.06,
  0.022)$ for all four replication projects.
  We can see that the two proportions agree fairly well for all values
  of $\alpha'$ considered. The number of discrepant studies in each project varies
  between 0 and 3.   Only in the Psychology project there are
  some studies which are successful with the replication success approach but 
  not the two-trials rule
  and some studies successful with the two-trials rule but not the replication 
  success approach.  The proportion of studies where both
  methods are successful (also shown in Figure \ref{fig:fig7}) is then smaller than the proportion of
  successful replications with either one of the two methods.  The
  three discrepant studies listed in the top three rows of
  Table~\ref{tbl:discrep} are an example of this particular feature.
}

\section{Discussion}\label{sec:discussion}
In this paper, we have expanded on the replication success approach introduced
in \citet{held2020} and demonstrated its advantages over alternative methods 
such as the two-trials rule. In particular, the method provides an attractive
compromise between hypothesis testing and estimation, as it penalizes 
shrinkage of the replication effect estimate compared to the original one,
while ensuring that both are statistically significant to some extent. 
For instance, the method will indicate only a low degree of replication success when 
the replication study shows a much smaller but statistically significant 
significant effect estimate, whereas it can still indicate a large degree of 
success when either original or replication $p$-value are slightly above the
significance level, provided their effect estimates are compatible.

We further refined the method by proposing the golden level, a
new threshold for replication success. It guarantees that borderline
significant original studies can only be replicated successfully if
the replication effect estimate is larger than the original
one. Compared to the two-trials rule, the golden level offers uniform
gains in project power and controls the Type-I error rate \hl{at any
  one-sided level $\alpha < 0.058$} if the
replication sample size is not \hl{smaller than the original one}. 
Empirical evaluation of data from four replication projects highlights
that in most cases the methods are in agreement, however, for the
study pairs where the approaches disagree, the replication success
approach seems to lead to more sensible conclusions. 
\hl{The good performance has been recently confirmed by a comparison of different
  replication success metrics through a simulation study in the
  presence of publication bias \citep{Muradchanian2020}.}

Despite a lack of agreement as to which statistical method should be used to 
evaluate replication studies, conclusions based on different methods usually 
agree. Nevertheless, in some cases, classical methods such as the two-trials 
rule may produce anomalies. We argue that 
the replication success approach improves upon existing methods leading to 
more appropriate inferences and decisions that better 
reflect the available evidence.
\hl{However, in extreme cases the performance of the sceptical $p$-value may be
  considered as strange or even counterintuitive. Specifically, if the
  original study was only borderline significant, a highly significant
  replication study can only lead to success if the replication effect
  estimate is larger than the original one. To understand this
  behaviour it is important to realize that the proposed approach does
  not synthesize the evidence from the two studies (like a standard
  meta-analysis). The sceptical $p$-value is designed to confirm
  claims of new discoveries through replication, but will remain
  ``stubborn'' \citep{LyWagenmakers2020} if the original study was not
  particularly convincing, even if there the replication study
  provides overwhelming evidence for an effect.  It will lead to a
  different result if the order of studies was reversed, as long as
  original and replication study do not have the same sample size
  ($c \neq 1$).  The related harmonic mean $\chi^2$-test
  \citep{held2020b} for evidence synthesis of two or more studies
 also requires  each study to be convincing on its own to a certain degree, but  
  treats them as exchangeable. }

With this paper we further advanced the reverse-Bayes methodology for
the analysis and design of replication studies, yet certain
limitations and opportunities for future research remain: First,
assuming normality of the effect estimates may be questionable,
especially for small sample sizes, and more robust distributional
assumptions could be considered.  Second, in some types of analyses
(\eg regression or ANOVA) the effect estimate is a vector and the
approach would need suitable adaptations. 
Third, there is a recent trend to not only conduct one but
several replications for one original study \citep[\eg][]{Klein2014, Ebersole2016, Klein2018}.
Also for this situation,
the method would need to be adapted, \eg the replication estimates
could be first synthesized and an analysis of replication
success could be performed subsequently.

\hl{Throughout the paper we have assumed that the relative sample size is
fixed in advance.  In practice the sample size of the replication
study is often chosen based on the result of the original study
\citep{Anderson2017}.  Power calculations as shown in Figure
\ref{fig:fig3} can then be inverted to determine the appropriate
sample size of the replication study. We can also invert 
equation \eqref{eq:res} to obtain the required replication sample size 
based on the specification of the minimum relative effect size $\dmin$
to achieve replication success. 
This novel way of calculating the sample size requires the
specification of the minimum relative effect size which can
still be considered as acceptable.  Sample size calculations based on
the two-trials rule can also be formulated in terms of the minimum
relative effect size by inverting equation \eqref{eq:dSig}.
We will report on a detailed comparison of the
different approaches in future work. }

\section*{Data and Software Availability} 
Data analyzed in this
article and software are available in the R-package \texttt{ReplicationSuccess},
which can be installed by running the following command in an R console: 
\texttt{install.packages("ReplicationSuccess", repos = \\"{http://R-Forge.R-project.org}"}).
Further information on
data preprocessing can be found on the corresponding help page 
(with the command \texttt{?RProjects}).

\section*{Acknowledgments}
Support by the Swiss
National Science Foundation (Project \#~189295) is gratefully
acknowledged. We acknowledge helpful and
constructive comments by the Editor and a referee on an earlier version of this article.

\bibliographystyle{imsart-nameyear} 
\bibliography{../antritt}


\end{document}